\documentclass[aps,prl,twocolumn,groupedaddress,showpacs]{revtex4-1}

\usepackage{graphicx,color}
\usepackage{epsfig}
\usepackage{dcolumn}
\usepackage{amsmath,amssymb,bm}
\usepackage{hyperref,url}
\usepackage[noabbrev]{cleveref}
\usepackage{CJK}
\usepackage{lineno}
\usepackage[english]{babel}
\usepackage[T1]{fontenc}
\usepackage{pifont}
\usepackage{amsfonts}
\usepackage{wasysym}
\usepackage{amsbsy}
\usepackage{psfrag}
\usepackage{color}
\usepackage{multirow}
\usepackage{cases}
\usepackage{stackengine}
\usepackage{float}
\usepackage{blkarray}
\usepackage{cancel}
\usepackage{mathrsfs}
\usepackage{empheq}
\usepackage{mathptmx} 

\begin{document}


\title{Axisymmetric bubble pinch-off at high Reynolds numbers}


\author{J.M. Gordillo}
\altaffiliation{Corresponding author: jgordill@us.es}
\affiliation{\'Area de Mec\'anica de Fluidos, Departamento de
Ingenier\'{\i}a Energ\'etica y Mec\'anica de Fluidos. Universidad de
Sevilla. \mbox{Avenida de los Descubrimientos s/n}, 41092, Sevilla,
Spain.}

\author{A. Sevilla}
\affiliation{\'Area de Mec\'anica de Fluidos, Departamento de Ingenier\'ia T\'ermica y de Fluidos. Universidad Carlos III de Madrid, Avenida de la Universidad 30, 28911, Legan\'es, Madrid, Spain.}

\author{J. Rodr\'iguez-Rodr\'iguez}
\affiliation{\'Area de Mec\'anica de Fluidos, Departamento de Ingenier\'ia T\'ermica y de Fluidos. Universidad Carlos III de Madrid, Avenida de la Universidad 30, 28911, Legan\'es, Madrid, Spain.}

\author{C. Mart\'inez-Baz\'an}
\affiliation{\'Area de Mec\'anica de Fluidos, Departamento de Ingenier\'ia T\'ermica y de Fluidos. Universidad Carlos III de Madrid, Avenida de la Universidad 30, 28911, Legan\'es, Madrid, Spain.}

\date{\today}

\begin{abstract}
Analytical considerations and potential flow numerical simulations of the pinch-off of bubbles at high Reynolds numbers reveal that the bubble minimum radius, $r_n$, decreases as $\tau\propto r_n^2 \, (-\ln{r_n^2})^{1/2}$, where $\tau$ is the time to break-up, when the local shape of the bubble near the singularity is symmetric. However, if the gas convective terms in the momentum equation become of the order of those of the liquid, the bubble shape is no longer symmetric and the evolution of the neck changes to a $r_n\propto\tau^{1/3}$ power law. These findings are verified experimentally.
\end{abstract}


\maketitle


The mechanisms underlying the capillary driven break-up of drops or liquid threads in a gas environment have been precisely described and experimentally verified during the last decade~\citep{Eggers97, Dayetal1998,Basaranexpnum, Eggers05}. However, despite its obvious relevance in many industrial applications, the generation and break-up of bubbles has received less attention in the literature. Therefore, the purpose of this Letter is to contribute to the understanding of the final stages of bubble pinch-off at high Reynolds numbers.

It is well known that, under inviscid conditions, the pinch-off dynamics of droplets is governed by a \emph{local} balance between surface tension forces and inertia (hereafter inertia will refer to the material derivative of linear momentum, $\rho \, D u/Dt$). Therefore, the near field structure of drop pinch-off is self-similar and \emph{universal}, in the sense that it does not depend either on initial nor on the far field flow conditions. Moreover,~\citet{Dayetal1998} and~\citet{LeppinenLister2003} reported that the interface had a self-similar, highly asymmetric double-cone shape at times close to the finite-time singularity.

On the other hand, recent experiments of bubble break-up in a highly viscous liquid \cite*{Bubpinchoff} have confirmed previous experimental and numerical results \cite*{Wendy, basaran04}, which clearly showed that the low Reynolds number break-up of bubbles is symmetric, and that the minimum radius approaches to zero linearly with $\tau$, where $\tau=t_b -t$  with $t_b$ being the break-up time. In addition, Suryo et al. \cite{basaran04} reported that the local shape of the interface near the singularity is parabolic and that the pinch-off is not self-similar. However, the case of bubble break-up in a low viscosity liquid (e.g. air in water), under study in the present work, seems to be slightly more subtle than its viscous counterpart. Indeed, Leppinen \& Lister \cite{LeppinenLister2003}  found that, under the assumption of potential flow, the self-similar solution near pinch-off is no longer stable when $\Lambda<1/ 6.2$, where $\Lambda=\hat{\rho}_g/\hat{\rho}_l$ is the inner (gas in our case) to outer (liquid in our case) density ratio. Moreover, in Leppinen \& Lister's study, the radius of the neck, $r_n$, behaves as $\tau^{2/3}$ for $\tau\rightarrow 0$. Nevertheless, recent accurate experimental measurements reported in \cite{Bubpinchoff}, show that $r_n$ behaves as $\tau^{1/2}$ for $\tau\rightarrow 0$, in agreement with previous analytical predictions \cite{Higgins}. The reason for the different exponent in the power law given in \cite{LeppinenLister2003}, and that of \cite{Higgins, Bubpinchoff}, lies on the fact that, whereas in the former study the pinch-off is promoted by surface tension, in the latter it is solely driven by the liquid inertia, given by $\hat{\rho_l} \, D \hat{u}_l/D\hat{t}$. Consequently, in this Letter we try to solve the apparent contradictions found by the above mentioned studies presenting both experimental evidences and potential flow numerical simulations similar to those reported in \cite{JFM05} and \cite{JFM05c} respectively, and we propose a slight correction to the above mentioned $r_n\propto \tau^{1/2}$ law.

Previous numerical results \cite{JFM05c} show that a bubble breaks-up symmetrically when it is placed at the stagnation point of a straining flow given by the following dimensionless velocity potential $\phi$ at infinity
\begin{equation}
\phi=-1/8\, r^2+1/4\,z^2,  
\label{potencial}
\end{equation}
where $r$ and $z$ are the dimensionless radial and axial cylindrical coordinates respectively. Here, distances, velocities and densities have been made dimensionless using $\hat{R}$ (initial bubble radius), $\hat{U}_l$ (characteristic outer flow velocity) and $\hat{\rho}_l$ respectively, and dimensional variables are indicated with a hat over them. The numerical method used to solve the coupled system of Laplace and Bernoulli equations that govern the gas flow inside the bubble and the outer liquid flow can be briefly described as follows. Provided the values of the inner and the outer potentials along the free surface at a given time, the velocities normal to the interface are computed through a boundary integral method that solves the discretized version of the both the liquid and the gas Green's integral equations. The values of the potentials are explicitly updated in time making use of a modified version of Bernoulli's equation that takes into account both the inner and the outer fluid densities, whereas the new positions of the nodal points are obtained by moving them normal to the interface. A more detailed description of the numerical method can be found in \cite{JFM05c}. With the aim at establishing the appropriate dependence of $r_n$ on $\tau$, we carried out numerical simulations of the evolution of the interface of a bubble initially located within the straining flow given by equation (\ref{potencial}) either at $r=0$, $z=z_o=0$ (\emph{symmetric} case) or at $r=0$, $z=z_o>0$ (\emph{asymmetric} cases). In our simulations we considered two different values of the inner to outer density ratio, namely $\Lambda=1.2\times 10^{-3}$ and $\Lambda=1.2\times 10^{-4}$.

\begin{figure}
\begin{center}
\includegraphics[width=0.4\textwidth]{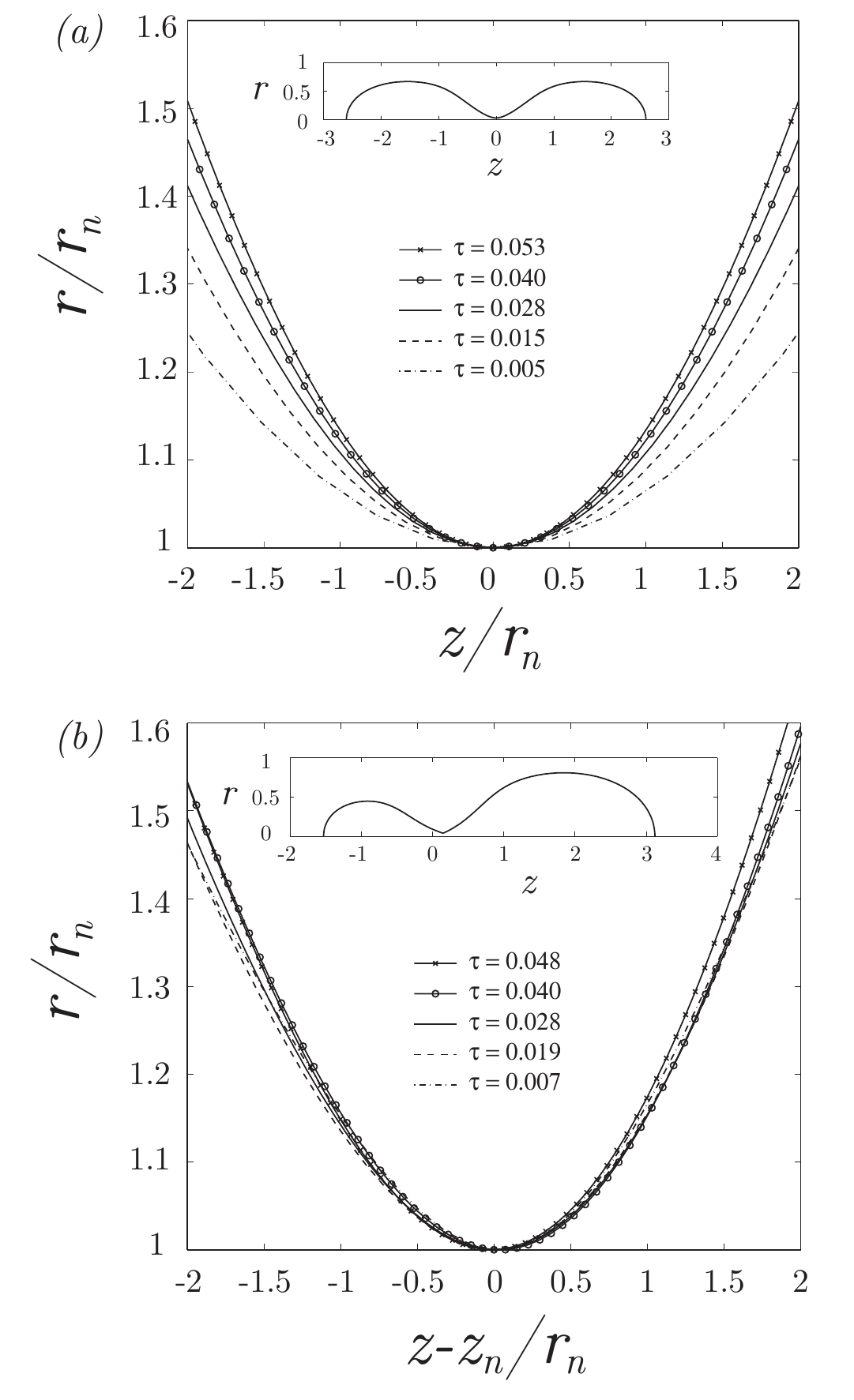}
\caption{Time evolution of the pinch-off region of a bubble, $\Lambda=1.2\times 10^{-3}$, at $We=\hat{\rho}_l\hat{U}_l^2\hat{R}/\hat{\sigma}=12$.  In both cases, the number of discrete points along the interface, 513, were rearranged at every time step to keep them equispaced. a) Symmetric break-up ($z_o=0$) and b) asymmetric case ($z_o=0.25$). Note that, while in 1a the radial scale is stretched more rapidly than the axial one, 1b suggests that a self-similar solution could be reached near pinch-off time if the KHI were prevented.}
\label{neckshapes}
\end{center}
\end{figure}

\begin{figure}
\includegraphics[width=0.5\textwidth]{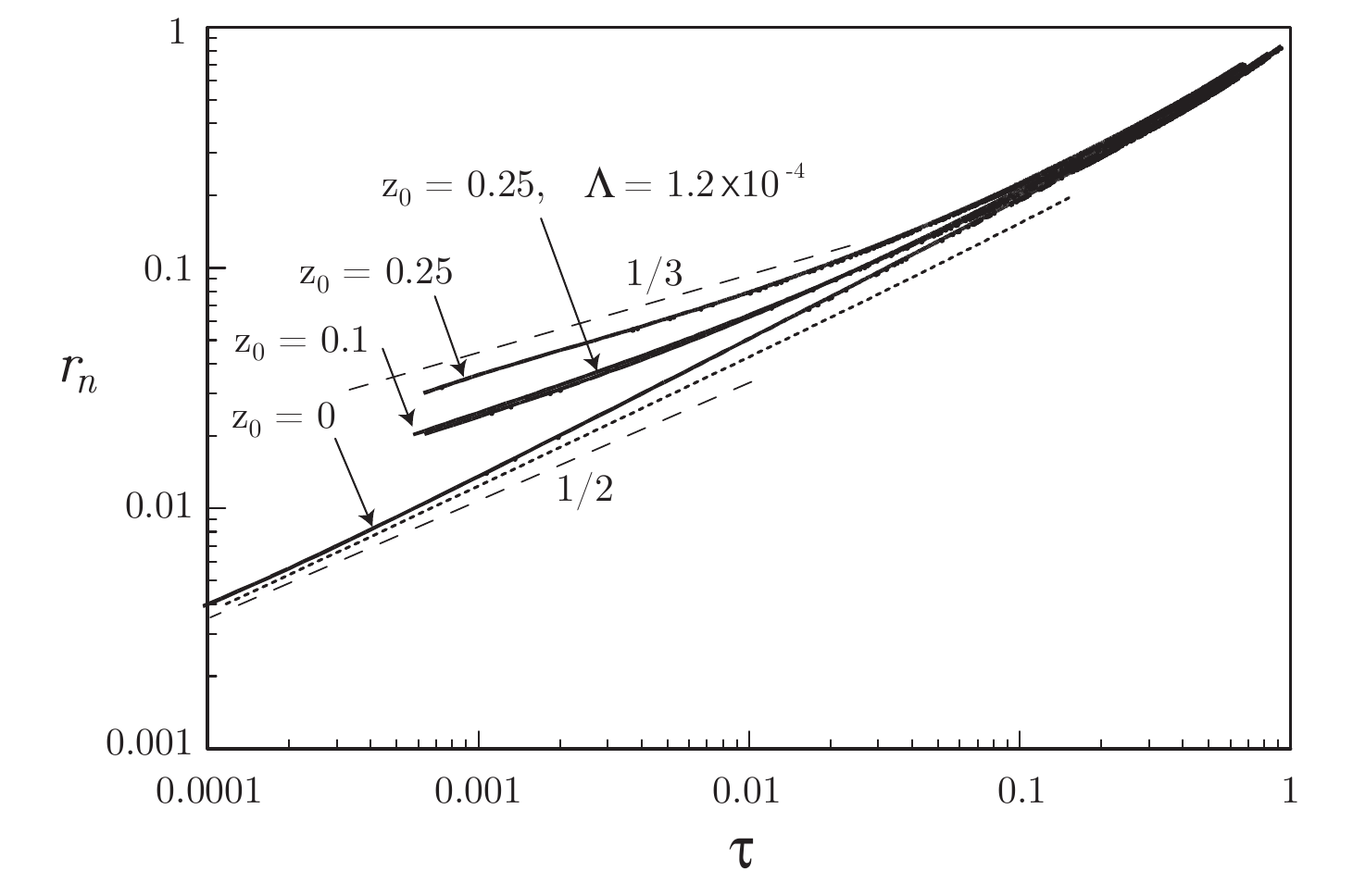}
\caption{Time evolution of the radius of the neck, $r_n(\tau)$, for different values of $z_o$ and $\Lambda$ (otherwise stated, $\Lambda=1.2\times 10^{-3}$) at $We=\hat{\rho}_l\hat{U}_l^2\hat{R}/\hat{\sigma}=12$. Notice that the transition from $1/2$ to the $1/3$  power law is delayed as $\Lambda$ and/or $Q_g$ decrease. The results given for $z_o=0.25$, $\Lambda=1.2\times 10^{-4}$ ($Q_g \simeq 0.59$, $q\simeq 0.197$, $r_{tran}\simeq 3.3\times 10^{-2}$) and those given for $z_o=0.1$, $\Lambda=1.2\times 10^{-3}$ ($Q_g\simeq 0.26$, $q\simeq 0.21$, $r_{tran}\simeq 4.3\times 10^{-2}$) are almost identical since the values of $r_{tran}$ are nearly the same in both cases. Dashed lines represent $\tau=2.2\times r_n^2\sqrt{-\ln{r_n^2}}$. \label{rneck}}
\end{figure}

Figure \ref{neckshapes} shows that, in the \emph{symmetric} case, the local bubble shape near the singularity is a parabola. It can be inferred from Fig.\ref{neckshapes}(a) that the radial length scale decreases more rapidly than the axial one and, consequently, the region near the singularity can be considered slender. Thus, to determine the proper dependence of $r_n$ on $\tau$, the analysis of the free-surface, potential-flow problem simplifies to solve the following ODE when $r_n\ll 1$,
\begin{equation}
\ln r_n(r'_n \, r_n)'+\frac{1}{2} \, \left( r'_n \right)^2=0.
\label{ode}
\end{equation}
Equation (\ref{ode}) can be integrated using the change of variables $q=r'_n\,r_n$ and $s=-\ln r_n$, where $q$ is simply a dimensionless flow-rate per unit length. A first integration of equation \ref{ode} yields $q\sqrt{s}=2^{-3/2}\,A$. Integrating again gives $A\tau=r_n^2 \, \sqrt{-\ln{r_n^2}}\times\left[1+O((\ln{r_n^2})^{-1})\right]$, indicating that $q\rightarrow 0$ as $\tau\rightarrow 0$. Notice that our analytical result provides a dependence of $r_n$ on $\tau$ with a slope, in a log-log scale, larger than 1/2, but which monotonically decreases towards  1/2 for $\ln r_n\rightarrow -\infty$. Figure \ref{rneck} shows how the numerical evolution matches the asymptotic behavior for $\tau < 0.01$, which slightly differs from the $r_n\propto \tau^{1/2}$ power law reported in \cite{Higgins, Bubpinchoff}. It needs to be pointed out that the deviation form the 1/2 power law has been previously experimentally measured at the Physics of Fluids group of Twente University. Finally, it should be pointed out that, since $\hat{\rho}_l\,\hat{q}/\hat{\mu}_l\rightarrow 0$ for $\tau\rightarrow 0$, the liquid viscosity must be retained to describe the latest instants prior to pinch-off.

The local behavior in the \emph{asymmetric} cases, displayed in Fig.\ref{neckshapes}(b), differs from its symmetric counterpart in the fact that the interface tends to form a double cone with different semiangles. Furthermore, Fig.\ref{rneck} shows that this type of break-up leads to $r_n\propto \tau^{1/3}$ at the latest instants of pinch-off.

\begin{figure}[t]
\includegraphics[width=0.5\textwidth]{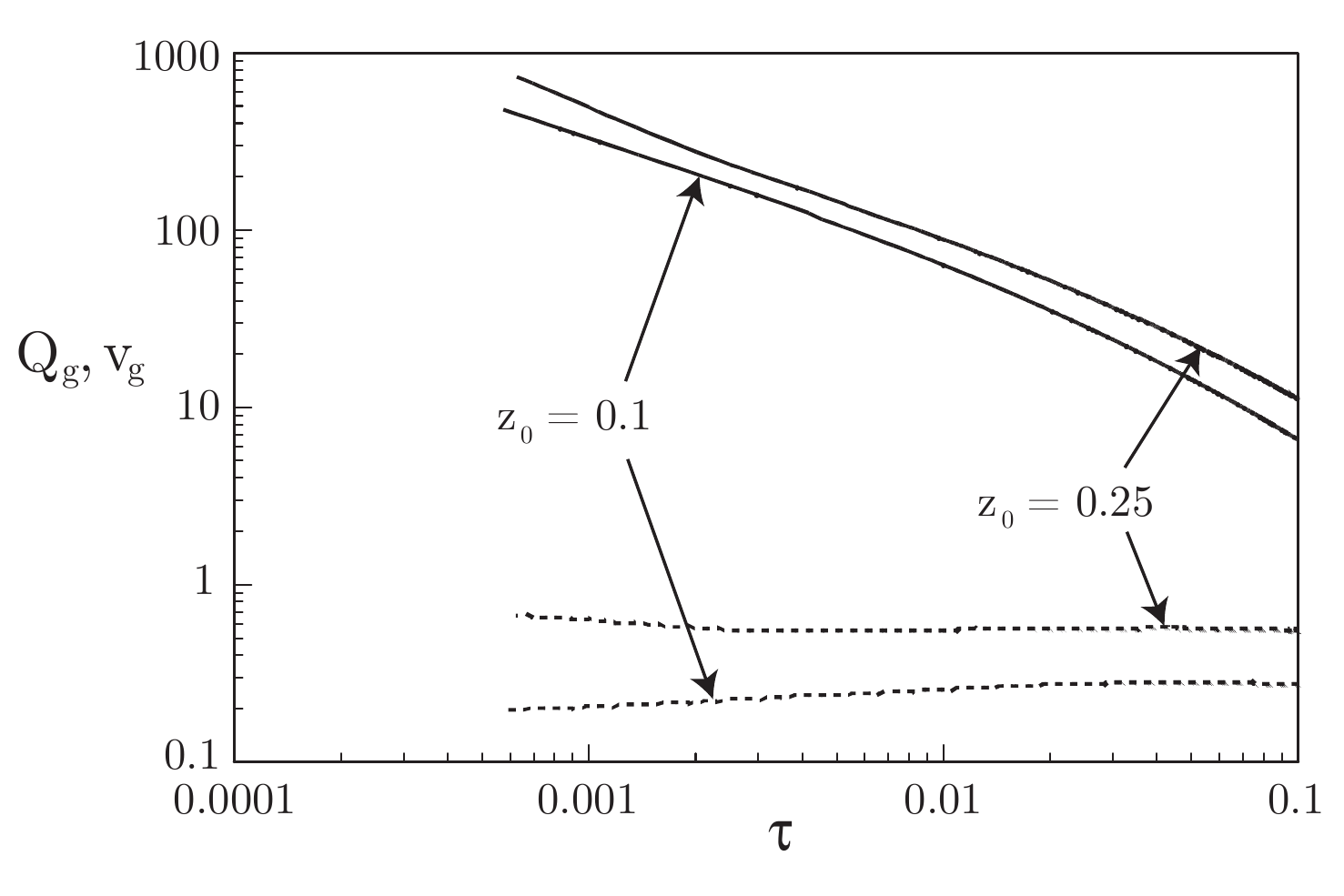}
\caption{Time evolution of $Q_g$ (dashed lines) and $v_g$ (solid lines) for $z_o=0.25, 0.1$, $\Lambda=1.2\times 10^{-3}$ and $We=12$. It can be observed that, whereas $v_g$ increases almost two decades, $Q_g$ remains nearly constant during the break-up process.}
\label{Qgasmin}
\end{figure}

The existence of the $1/3$ power law can be physically explained as follows. Whereas the symmetry imposes a zero gas velocity inside the neck ($v_g=0$), in the asymmetric break-up $v_g$ increases as $\tau\rightarrow 0$, as shown in Fig.\ref{Qgasmin}. Thus, as the gas flows through the neck at a velocity $v_g$, there is a suction originated by the high speed gas stream (Bernoulli's effect), which further accelerates the outer liquid towards the axis, explaining the faster asymmetric collapse. As pointed out by Leppinen \& Lister \cite{LeppinenLister2003}, such mechanism is responsible for the unstable behavior near the singularity for $\Lambda<6.2^{-1}$. In fact, this Kelvin-Helmholtz instability (KHI) was also obtained in our numerical simulations, which were stopped as soon as there was a strong interface distortion near the minimum radius.

In the following we will show that the $1/3$ power law appears because the final stages of pinch-off are not driven either by surface tension nor by liquid inertia solely, but by both liquid and gas inertia. First, note that dimensional arguments require a typical length to determine the characteristic gas velocity in our potential flow problem which cannot be the capillary one since surface tension effects are negligible in a collapse following a 1/3 power law ($We_{n}=\hat{\rho}_l\,(d\hat{r}_n/d\hat{\tau})^2\,\hat{r}_n/\hat{\sigma}\propto \tau^{-1}$, $\hat{\sigma}$ being the liquid-gas surface tension). The solution to this apparent paradox is given by a quick inspection of Fig.\ref{Qgasmin} where it is shown that, in our incompressible, inviscid simulations, and within admissible numerical errors, $v_{g}\times r_n^2=Q_g\simeq constant$, where $Q_g$ stands for the dimensionless gas flow-rate. Thus, the balance between the liquid inertia, $\hat{\rho}_l D\hat{u}_l/Dt$, and the pressure drop in the gas stream requires that $\hat{\rho}_l\,\hat{r}_n/\hat{\tau}^2\sim \hat{\rho}_g\,\hat{Q}_g^2/\hat{r}_n^5$, indicating that $r_n$ and $v_g$ can be appropriately defined as $r_n \sim \Lambda^{1/6}\,Q_g^{1/3}\,\tau^{1/3}$ and $v_g \sim \Lambda^{-1/3}\,Q_g^{1/3}\,\tau^{-2/3}$ respectively. Note that, in the above discussion of the asymmetric case ($Q_g\neq 0$), we have considered the gas flow to be quasi-steady since the gas residence time, $\hat{t}_r\sim \hat{r}_n^3/\hat{Q}_g$ is much smaller than the characteristic time of the problem, $\hat{t}_0$, which can be either $\hat{t}_0\sim\,\hat{r}^2_n/\hat{q}$ or $\hat{t}_0\sim\,\hat{r}_n^3/\hat{Q}_g\,\Lambda^{-1/2}$ and thus, $t_r/t_0\sim r_n\,(q Q_g)\ll 1$ or $t_r/t_0\sim \Lambda^{1/2}\ll 1$. Also notice that the transition to the $1/3$ power law will take place at a characteristic transition radius $r_{tran}$ such that the liquid and the gas inertia are both of the same order of magnitude, $\hat{\rho}_l\,(\hat{q}/\hat{r}_{tran})^2\sim\,\hat{\rho}_g\,(\hat{Q}_g/\hat{r}_{tran}^2)^2\rightarrow r_{tran}\sim\,\Lambda^{1/2}\,(Q_g/q(\tau))$. Clearly, the smaller $Q_g/q(\tau)$ is for $\tau\sim O(1)$ (or, equivalently, the smaller the initial asymmetry), the smaller is the transition radius and the more difficult will be to determine, either numerically or experimentally, the existence of the $1/3$ power law. This conclusion is supported by Fig.\ref{rneck}, where it is shown that, under the potential flow approximation, this transition exists whenever there is an initial asymmetry ($z_o =0.10$ and $z_o =0.25$). Consistently with the above discussion, in the symmetric case no transition is observed within the spatial resolution of our numerical method. Furthermore, the power law transition is delayed as $\Lambda$ or $Q_g$ decrease (see Figs.  2 and 3). Figure 2 also shows that, in the case of $z_o =0.25$, the transition radius is smaller for $\Lambda=1.2 \times 10^{-4}$ than that obtained for  $\Lambda=1.2 \times 10^{-3}$. It should also be emphasized that neither $Q_g$ nor $q$ are fixed locally, but through the entire flow domain.

So far we have assumed that the characteristic axial length scale near the singularity is the same as the characteristic radial length scale, i.e. $r_n$. However, as already discussed above, the strong shear near the minimum radius favors the development of a KHI, for which all the lengths greater than $\delta_m$, with $\delta_m\sim (Q_g^2\,We\,\Lambda)^{-1}\,r_n^4\ll r_n$ being the characteristic dimensionless length scale such that surface tension and gas inertia are both of the same order of magnitude, will be unstable. Here, $We=\hat{\rho}_l\hat{U}_l^2\hat{R}/\hat{\sigma}$ and, consequently, if the latest instants previous to break-up could be accurately simulated, dendrites of a typical length scale $\delta_m\ll r_n$, analogous to the ones identified in \cite{LeppinenLister2003}, would develop.

The applicability of the above results to real flows depends on the validity of the potential flow approximation employed here.
Although the thickness of the gas viscous boundary layer $\delta_g\sim\, r_n\,Re_g^{-1/2}\propto r_n^{3/2}$, where $Re_g=\hat{\rho}_g\,\hat{Q}_g/(\hat{\mu}_g\,\hat{r}_n)$, is very small compared with the radius of the neck, $\delta_g\ll r_n$, flow
separation may take place downstream from the axial location of $r_n$. Indeed, in a first approach, the gas flow downstream of
the neck is quasi-steady and similar to that of a divergent diffuser with some blowing at the walls, whose opening semi-angle is larger than the separation one. Under these conditions the flow may separate avoiding the gas stagnation pressure to fully recover
inside the growing bubble. Furthermore, the flow separation will decrease the above mentioned suction effect and, consequently, $\hat{Q}_g$ will also decrease, producing a subsequent accumulation of gas upstream of the neck. To conclude, the suction mechanism, which accelerates the outer liquid radially toward the axis, will be less effective in real flows than it is predicted by our simulations. In addition, viscous effects in the liquid are expected to be negligible since $Re_l\propto\tau^{-1/3}$.

\begin{figure}[t]
\begin{center}
\includegraphics[width=0.5\textwidth]{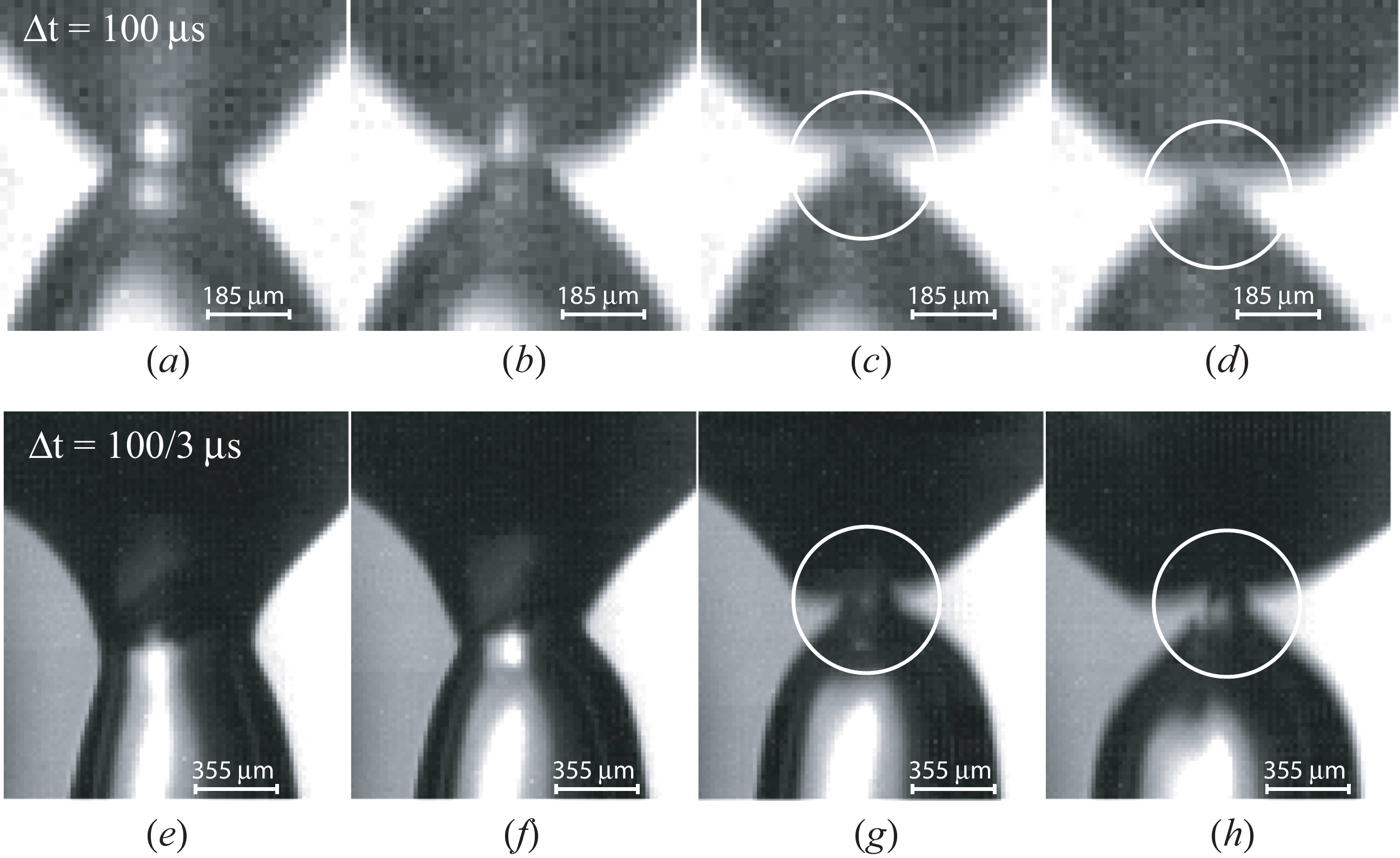}
\caption{Close-up views showing two showing two different sequences of the bubble pinch-off corresponding to the experimetal conditions of Fig.\ref{Fig4}. The air and water flow rates are $Q_a$= 610 ml/min, $Q_w$= 4 l/min in photographs (a)-(d), and $Q_a$= 850 ml/min, $Q_w$= 6 l/min in photographs (e)-(h). The time interval between the shown frames is $\Delta t$= 100 $\mu s$ in (a)-(d), and $\Delta t$= 100/3 $\mu s$ in (e)-(h). The spatial resolution is, in all cases, 18.6 $\mu m/px$, and the movies were recorded with a shutter time of 1/30000 $s^{-1}$. Gas and liquid Reynolds numbers based on the needle diameter are, for both experimental conditions, of the order of $\sim O(10^3)$ and  $\sim O(10^4)$ respectively. \label{Fig5}}
\end{center}
\end{figure}

Up to now we have presented our numerical results to describe the time evolution of the neck during the bubble break-up process. However, in addition to the numerical simulations  we have also performed a meticulous series of experiments of bubble formation in a co-flowing liquid stream using the experimental facility described in \cite{JFM05}. To study the final stages of the bubble break-up, we used a high-speed video camera to record sequences of the break-up of bubbles at a rate which was varied between $3\times 10^4$ fps (resolution of $256\times 128$ pixels) and $5\times 10^4$ fps (resolution of $256\times 64$ pixels) depending on the experimental conditions.

\begin{figure}[]
\begin{center}
\includegraphics[width=0.5\textwidth]{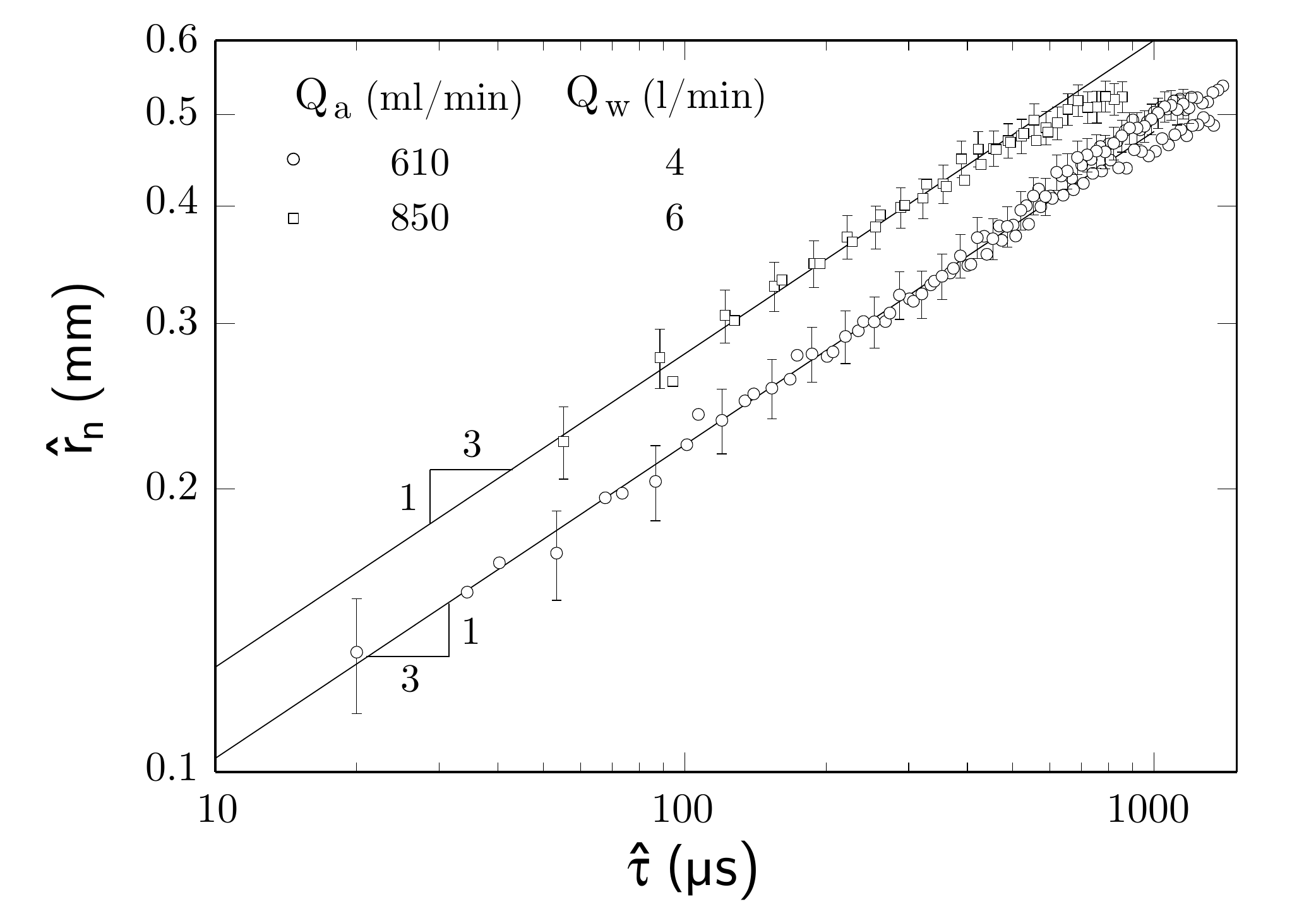}
\caption{Experimental results for the evolution of $\hat{r}_n$ with $\hat{\tau}$ in a co-flowing air-water jet. The gas is injected through a needle (inner and outer radii $0.419$ and $0.635$ mm respectively) placed coaxially within a liquid jet discharging from a $4$ mm circular nozzle. $Q_a$ and $Q_w$ respectively denote the air and water flow rates. A detailed description of the experimental facility is given in \cite{JFM05}.}
\label{Fig4}
\end{center}
\end{figure}

An example of two different sequences is shown in Fig.\ref{Fig5}. Before we proceed any further, we would like to mention that the experiments reported here were previously selected to make sure that  break-up was purely axisymmetric. Figure \ref{Fig4} shows the neck radius evolution as it approaches the singularity (pinch-off point). This figure reports an experimental evidence of the $r_n\sim \tau^{1/3}$ power law in spite of the flow separation. As already discussed, the clear departure from the $1/2$ power law is promoted by the favorable (negative) pressure gradient induced by the gas-flow towards the neck. The break-up time was determined from the recordings by least-squares fitting of the last 6 to 10 data points to the 1/3 power law, since fitting to the $\tau\propto r_n^2\sqrt{-\ln{r_n^2}}$ or to the 2/3 power law resulted in inconsistent results for the pinch-off time when compared to the high-speed movies. It was checked that the results were not sensitive to the number of selected data points and, in all cases, the correlation coefficient was very close to unity.

More importantly, the photographs shown in Fig.\ref{Fig5} provide evidence of the asymmetry generated by the favorable pressure gradient induced by the gas flow in the neck region, which manifests in the form of an entrainment of liquid within the forming bubble (see frames (c)-(d) and (g)-(h) in Fig.\ref{Fig5}).

In view of the previous results, it can be concluded that, in spite of flow separation, the favorable pressure gradient generated within the neck region accelerates the liquid both radially and axially, producing a strong asymmetry near the neck region. Since this mechanism is only appreciable at scales $r_{tran}\sim\,\Lambda^{1/2}\,(Q_g/q)$, detailed experiments need to be carried out in order to verify whether this behavior, experimentally verified for large values of $Q_g/q$, is also present if, for $\tau\sim O(1)$, $Q_g/q(\tau)\lesssim O(1)$.

\begin{acknowledgments}
The authors are greatful to the PoF group at Twenty University, Enschede, The Netherlands, and specially to Prof. Detlef Lohse, for their hospitality and for sharing with us their experimental results. We also wish to thank the Department of Continuum Mechanics and Structural Analysis, UC3M, for sharing their high-speed camera with us.This research has been supported by the Spanish MCyT under Projects No. DPI2002-04550-C07-06 and AGL2000-0374-P4-02.
\end{acknowledgments}


\begin{thebibliography}{11}%
\makeatletter
\providecommand \@ifxundefined [1]{%
 \@ifx{#1\undefined}
}%
\providecommand \@ifnum [1]{%
 \ifnum #1\expandafter \@firstoftwo
 \else \expandafter \@secondoftwo
 \fi
}%
\providecommand \@ifx [1]{%
 \ifx #1\expandafter \@firstoftwo
 \else \expandafter \@secondoftwo
 \fi
}%
\providecommand \natexlab [1]{#1}%
\providecommand \enquote  [1]{``#1''}%
\providecommand \bibnamefont  [1]{#1}%
\providecommand \bibfnamefont [1]{#1}%
\providecommand \citenamefont [1]{#1}%
\providecommand \href@noop [0]{\@secondoftwo}%
\providecommand \href [0]{\begingroup \@sanitize@url \@href}%
\providecommand \@href[1]{\@@startlink{#1}\@@href}%
\providecommand \@@href[1]{\endgroup#1\@@endlink}%
\providecommand \@sanitize@url [0]{\catcode `\\12\catcode `\$12\catcode
  `\&12\catcode `\#12\catcode `\^12\catcode `\_12\catcode `\%12\relax}%
\providecommand \@@startlink[1]{}%
\providecommand \@@endlink[0]{}%
\providecommand \url  [0]{\begingroup\@sanitize@url \@url }%
\providecommand \@url [1]{\endgroup\@href {#1}{\urlprefix }}%
\providecommand \urlprefix  [0]{URL }%
\providecommand \Eprint [0]{\href }%
\providecommand \doibase [0]{http://dx.doi.org/}%
\providecommand \selectlanguage [0]{\@gobble}%
\providecommand \bibinfo  [0]{\@secondoftwo}%
\providecommand \bibfield  [0]{\@secondoftwo}%
\providecommand \translation [1]{[#1]}%
\providecommand \BibitemOpen [0]{}%
\providecommand \bibitemStop [0]{}%
\providecommand \bibitemNoStop [0]{.\EOS\space}%
\providecommand \EOS [0]{\spacefactor3000\relax}%
\providecommand \BibitemShut  [1]{\csname bibitem#1\endcsname}%
\let\auto@bib@innerbib\@empty
\bibitem [{\citenamefont {Eggers}(1997)}]{Eggers97}%
  \BibitemOpen
  \bibfield  {author} {\bibinfo {author} {\bibfnamefont {J.}~\bibnamefont
  {Eggers}},\ }\href {\doibase 10.1103/RevModPhys.69.865} {\bibfield  {journal}
  {\bibinfo  {journal} {Rev. Mod. Phys.}\ }\textbf {\bibinfo {volume} {69}},\
  \bibinfo {pages} {865} (\bibinfo {year} {1997})}\BibitemShut {NoStop}%
\bibitem [{\citenamefont {Day}\ \emph {et~al.}(1998)\citenamefont {Day},
  \citenamefont {Hinch},\ and\ \citenamefont {Lister}}]{Dayetal1998}%
  \BibitemOpen
  \bibfield  {author} {\bibinfo {author} {\bibfnamefont {R.~F.}\ \bibnamefont
  {Day}}, \bibinfo {author} {\bibfnamefont {E.~J.}\ \bibnamefont {Hinch}}, \
  and\ \bibinfo {author} {\bibfnamefont {J.~R.}\ \bibnamefont {Lister}},\
  }\href {\doibase 10.1103/PhysRevLett.80.704} {\bibfield  {journal} {\bibinfo
  {journal} {Phys. Rev. Lett.}\ }\textbf {\bibinfo {volume} {80}},\ \bibinfo
  {pages} {704} (\bibinfo {year} {1998})}\BibitemShut {NoStop}%
\bibitem [{\citenamefont {Chen}\ \emph {et~al.}(2002)\citenamefont {Chen},
  \citenamefont {Notz},\ and\ \citenamefont {Basaran}}]{Basaranexpnum}%
  \BibitemOpen
  \bibfield  {author} {\bibinfo {author} {\bibfnamefont {A.~U.}\ \bibnamefont
  {Chen}}, \bibinfo {author} {\bibfnamefont {P.~K.}\ \bibnamefont {Notz}}, \
  and\ \bibinfo {author} {\bibfnamefont {O.~A.}\ \bibnamefont {Basaran}},\
  }\href {\doibase 10.1103/PhysRevLett.88.174501} {\bibfield  {journal}
  {\bibinfo  {journal} {Phys. Rev. Lett.}\ }\textbf {\bibinfo {volume} {88}},\
  \bibinfo {pages} {174501} (\bibinfo {year} {2002})}\BibitemShut {NoStop}%
\bibitem [{\citenamefont {Eggers}(2005)}]{Eggers05}%
  \BibitemOpen
  \bibfield  {author} {\bibinfo {author} {\bibfnamefont {J.}~\bibnamefont
  {Eggers}},\ }\href {\doibase 10.1002/zamm.200410193} {\bibfield  {journal}
  {\bibinfo  {journal} {ZAMM}\ }\textbf {\bibinfo {volume} {85}},\ \bibinfo
  {pages} {400} (\bibinfo {year} {2005})}\BibitemShut {NoStop}%
\bibitem [{\citenamefont {Leppinen}\ and\ \citenamefont
  {Lister}(2003)}]{LeppinenLister2003}%
  \BibitemOpen
  \bibfield  {author} {\bibinfo {author} {\bibfnamefont {D.}~\bibnamefont
  {Leppinen}}\ and\ \bibinfo {author} {\bibfnamefont {J.~R.}\ \bibnamefont
  {Lister}},\ }\href {\doibase 10.1063/1.1537237} {\bibfield  {journal}
  {\bibinfo  {journal} {Phys. Fluids}\ }\textbf {\bibinfo {volume} {15}},\
  \bibinfo {pages} {568} (\bibinfo {year} {2003})}\BibitemShut {NoStop}%
\bibitem [{\citenamefont {Burton}\ \emph {et~al.}(2005)\citenamefont {Burton},
  \citenamefont {Waldrep},\ and\ \citenamefont {Taborek}}]{Bubpinchoff}%
  \BibitemOpen
  \bibfield  {author} {\bibinfo {author} {\bibfnamefont {J.}~\bibnamefont
  {Burton}}, \bibinfo {author} {\bibfnamefont {R.}~\bibnamefont {Waldrep}}, \
  and\ \bibinfo {author} {\bibfnamefont {P.}~\bibnamefont {Taborek}},\ }\href
  {\doibase 10.1103/PhysRevLett.94.184502} {\bibfield  {journal} {\bibinfo
  {journal} {Phys. Rev. Lett.}\ }\textbf {\bibinfo {volume} {94}},\ \bibinfo
  {pages} {184502} (\bibinfo {year} {2005})}\BibitemShut {NoStop}%
\bibitem [{\citenamefont {Doshi}\ \emph {et~al.}(2003)\citenamefont {Doshi},
  \citenamefont {Cohen}, \citenamefont {Zhang}, \citenamefont {Siegel},
  \citenamefont {Howell}, \citenamefont {Basaran},\ and\ \citenamefont
  {Nagel}}]{Wendy}%
  \BibitemOpen
  \bibfield  {author} {\bibinfo {author} {\bibfnamefont {P.}~\bibnamefont
  {Doshi}}, \bibinfo {author} {\bibfnamefont {I.}~\bibnamefont {Cohen}},
  \bibinfo {author} {\bibfnamefont {W.~W.}\ \bibnamefont {Zhang}}, \bibinfo
  {author} {\bibfnamefont {M.}~\bibnamefont {Siegel}}, \bibinfo {author}
  {\bibfnamefont {P.}~\bibnamefont {Howell}}, \bibinfo {author} {\bibfnamefont
  {O.~A.}\ \bibnamefont {Basaran}}, \ and\ \bibinfo {author} {\bibfnamefont
  {S.~R.}\ \bibnamefont {Nagel}},\ }\href {\doibase 10.1126/science.1089272}
  {\bibfield  {journal} {\bibinfo  {journal} {Science}\ }\textbf {\bibinfo
  {volume} {302}},\ \bibinfo {pages} {1185} (\bibinfo {year}
  {2003})}\BibitemShut {NoStop}%
\bibitem [{\citenamefont {Suryo}\ \emph {et~al.}(2004)\citenamefont {Suryo},
  \citenamefont {Doshi},\ and\ \citenamefont {Basaran}}]{basaran04}%
  \BibitemOpen
  \bibfield  {author} {\bibinfo {author} {\bibfnamefont {R.}~\bibnamefont
  {Suryo}}, \bibinfo {author} {\bibfnamefont {P.}~\bibnamefont {Doshi}}, \ and\
  \bibinfo {author} {\bibfnamefont {O.~A.}\ \bibnamefont {Basaran}},\ }\href
  {\doibase 10.1063/1.1793631} {\bibfield  {journal} {\bibinfo  {journal}
  {Phys. Fluids}\ }\textbf {\bibinfo {volume} {16}},\ \bibinfo {pages} {4177}
  (\bibinfo {year} {2004})}\BibitemShut {NoStop}%
\bibitem [{\citenamefont {Longuet-Higgins}\ \emph {et~al.}(1991)\citenamefont
  {Longuet-Higgins}, \citenamefont {Kerman},\ and\ \citenamefont
  {Lunde}}]{Higgins}%
  \BibitemOpen
  \bibfield  {author} {\bibinfo {author} {\bibfnamefont {M.~S.}\ \bibnamefont
  {Longuet-Higgins}}, \bibinfo {author} {\bibfnamefont {B.~R.}\ \bibnamefont
  {Kerman}}, \ and\ \bibinfo {author} {\bibfnamefont {K.}~\bibnamefont
  {Lunde}},\ }\href {\doibase 10.1017/S0022112091000836} {\bibfield  {journal}
  {\bibinfo  {journal} {J. Fluid Mech.}\ }\textbf {\bibinfo {volume} {230}},\
  \bibinfo {pages} {365} (\bibinfo {year} {1991})}\BibitemShut {NoStop}%
\bibitem [{\citenamefont {Sevilla}\ \emph {et~al.}(2005)\citenamefont
  {Sevilla}, \citenamefont {Gordillo},\ and\ \citenamefont
  {Mart\'{\i}nez-Baz\'an}}]{JFM05}%
  \BibitemOpen
  \bibfield  {author} {\bibinfo {author} {\bibfnamefont {A.}~\bibnamefont
  {Sevilla}}, \bibinfo {author} {\bibfnamefont {J.~M.}\ \bibnamefont
  {Gordillo}}, \ and\ \bibinfo {author} {\bibfnamefont {C.}~\bibnamefont
  {Mart\'{\i}nez-Baz\'an}},\ }\href {\doibase 10.1017/S002211200500354X}
  {\bibfield  {journal} {\bibinfo  {journal} {J. Fluid Mech.}\ }\textbf
  {\bibinfo {volume} {530}},\ \bibinfo {pages} {181} (\bibinfo {year}
  {2005})}\BibitemShut {NoStop}%
\bibitem [{\citenamefont {Rodr\'iguez-Rodr\'iguez}\ \emph
  {et~al.}(2006)\citenamefont {Rodr\'iguez-Rodr\'iguez}, \citenamefont
  {Gordillo},\ and\ \citenamefont {Mart\'{\i}nez-Baz\'an}}]{JFM05c}%
  \BibitemOpen
  \bibfield  {author} {\bibinfo {author} {\bibfnamefont {J.}~\bibnamefont
  {Rodr\'iguez-Rodr\'iguez}}, \bibinfo {author} {\bibfnamefont {J.~M.}\
  \bibnamefont {Gordillo}}, \ and\ \bibinfo {author} {\bibfnamefont
  {C.}~\bibnamefont {Mart\'{\i}nez-Baz\'an}},\ }\href {\doibase
  10.1017/S002211200500741X} {\bibfield  {journal} {\bibinfo  {journal} {J.
  Fluid Mech.}\ }\textbf {\bibinfo {volume} {548}},\ \bibinfo {pages} {69}
  (\bibinfo {year} {2006})}\BibitemShut {NoStop}%
\end{thebibliography}
\end{document}